# Virtually Bare Nanocrystal Surfaces – Significantly Enhanced Electrical Transport in $CuInSe_2$ and $CuIn_{1-x}Ga_xSe_2$ Thin Films upon Ligand Exchange with Thermally Degradable 1-Ethyl-5-thiotetrazole

*Jannika Lauth, Jakob Marbach, Andreas Meyer, Sedat Dogan,*
*Christian Klinke, Andreas Kornowski, and Horst Weller**

Institute of Physical Chemistry, University of Hamburg,
Grindelallee 117, D-20146 Hamburg, Germany

We present a facile and safe ligand exchange method for readily synthesized $CuInSe_2$ (CIS) and $CuIn_{1-x}Ga_xSe_2$ (CIGS) nanocrystals (NCs) from oleylamine to 1-ethyl-5-thiotetrazole which preserves the colloidal stability of the chalcopyrite structure. 1-ethyl-5-thiotetrazole as thermally degradable ligand is adapted for the first time for trigonal pyramidal CIS NCs (18 nm), elongated CIS NCs (9 nm) and CIGS NCs (6 nm). The exchanged NC solutions are spin-coated onto $Si/SiO_2$ substrates with predefined gold electrodes to yield ordered NC thin films. These films are thermally annealed at 260 °C to completely remove 1-ethyl-5-thiotetrazol leaving virtually bare NC surfaces. We measure the current-voltage characteristics of the NC solids prior to ligand thermolysis in the dark and under illumination and after thermolysis of the ligand in the same manner. The conductivity of trigonal pyramidal CIS NCs increases by four orders of magnitude from $1.4 \times 10^{-9}$ S/cm in the dark to $1.4 \times 10^{-5}$ S/cm for ligand-free illuminated NC films. Elongated CIS NC films show an increase by three orders of magnitude and CIGS NC films exhibit improved conductivity by two orders of magnitude. The degree of conductivity enhancement thereby depends on the NC size accentuating the role of trap-states and internal grain boundaries in ligand-free NC solids for electrical transport. Our approach offers for the first time the possibility to address chalcopyrite materials' electrical properties in a virtually ligand-free state.

## 1. Introduction

Chalcopyrite materials, especially $CuInSe_2$ (CIS) and $CuIn_{1-x}Ga_xSe_2$ (CIGS) represent promising candidates for regenerative energy conversion. The direct band gap of 1.04 eV for pure CIS can be adjusted by the amount of gallium incorporated into the crystal lattice resulting in an increased band gap of up to 1.15 eV which is close to the optimum of best sun conversion efficiencies at AM1.5.[1,2] This tunability and the high absorption coefficient surpassing $10^5$ $cm^{-1}$ predestine the material for the use in solar cells.[3] A controlled stoichiometry of the system is thereby essential and can be for example sufficiently achieved by wet-chemical colloidal synthesis approaches for CIS and CIGS nanocrystals (NCs).[4,5] The significant progress of NC materials becomes apparent in their still emerging application in multiple areas like light emitting diodes (LEDs)[6,7], lasers[8], field effect transistors (FETs)[7,9] and optoelectronics[4,10–14] in the last two decades. Especially the solar cell sector is an important field that profits from ongoing focused research conducted in the NC sector, given the fact, that closely packed monolayers of PbS and CdSe quantum dots can enhance broad-band absorption via dipolar coupling[15] and multijunction quantum dot solar cells can significantly surpass the Shockley-Queisser limit.[16,17]

A crucial drawback still preventing most NC species from being accessible and profitable for electronic applications is the way they are processed. Wet-chemical approaches on the one hand represent low-cost ways to synthesize the material but on the other hand the obtained NCs are mostly stabilized with long-chained organic capping agents that secure stability in solution and prevent aggregation. At the same time, these organic surfactant molecules inhibit electronic interparticle interactions resulting in insulating behavior of NC assemblies.[7]

There are ways to increase the conductivity in NC solids for example by using inorganic metal chalcogenide complex ligands (MCCs)[18,19] but these involve the use of toxic and hazardous anhydrous hydrazine and are hard to apply in industrial large-scale dimensions. Chemical treatments with 1,2-ethanedithiol[20] and ethanethiol[21] show that photovoltaic device efficiency and electrical transport of NC films can be enhanced but ligand exchange is performed after the deposition of the NCs confining the flexibility of device fabrication. Further approaches like the use of thiocyanate as capping agent[22] and Meerwein's salt[23] for removing stabilizing amines, carboxylates and phosphonates have not yet been shown for CIS and CIGS. Recent results in stripping ligands off of NC surfaces with $(NH_4)_2S$[24] may be promising but have to show their abilities in determining the electrical properties of the NC materials.

Here, we report the characterization of two chalcopyrite NC systems CIS and CIGS with regard to their electrical transport properties once the initial oleylamine (OLA) ligand shell is exchanged by 1-ethyl-5-thiotetrazole (ETT). The capability of ETT as ligand and reactant for CdS NCs has been described by Voitekhovich et al.[25] Here it is successfully implemented for the first time for CIS and CIGS NCs synthesized by modified methods of Panthani et al.[4] and Koo et al.[26] Our surfactant exchange method offers the possibility to control NC film thickness by the number of spin-coating steps. ETT as ligand represents a safe and straightforward opportunity to obtain highly conductive organics-free CIS and CIGS NC solids that have great potential in photovoltaic applications.

## 2. Results and Discussion

### 2.1. Characterization of CIS and CIGS NCs

CIS and CIGS NCs were characterized using Transmission Electron Microscopy (TEM) and powder X-Ray Diffraction (XRD). Trigonal pyramidal CIS NCs with an average size of 19 nm (± 6 nm, edge length) were synthesized according to Koo et al.[26] (see Experimental for detailed syntheses and Supporting Information for XRD of the NCs). The TEM image in **Figure 1**a depicts the results of a typical ligand exchanged NC solution of trigonal pyramidal CIS. The structure and size distribution of the NCs remain unaltered after ligand exchange. (TEM images prior to ligand exchange, Figure S1, corresponding XRD Figure S2). Elongated

CIS NCs with an average size of 9 nm (± 2 nm) and CIGS NCs with a mean diameter of 6 nm (± 1 nm) were obtained by modifying a Panthani et al.[4] method and likewise ligand exchanged with ETT, see Figure 1b and 1c (TEM images prior to ligand exchange, Figure S3, (XRD Figure S4) and S5 (XRD Figure S6)).

**2.2. Ligand Exchange of CIS and CIGS NCs with ETT**

*2.2.1. Ligand Exchange Procedure*

Ligand exchange of the NCs with ETT was conducted in solution. In a typical experiment 1 mL of a 0.025 M trigonal pyramidal CIS NC solution in chloroform, that had been precipitated once with 2-propanol, was mixed with 2 mL of a 2.8 M ETT chloroform solution (220 fold molar ligand excess, conditions for ligand exchange, see Table S1). The ligand exchange mixture was stirred at 50 °C for 20 respectively 30 hours for CIGS NCs and subsequently precipitated once with *n*-hexane. CIS and CIGS NCs maintained their size distribution and appearance (Figure 1).

*2.2.2. Attenuated Total Reflectance Fourier Transform Infrared Spectroscopy (ATR-FTIR)*

Ligand exchange reactions were monitored using Attenuated Total Reflectance Fourier Transform Infrared Spectroscopy (ATR-FTIR). **Figure 2** shows IR spectra of initial OLA capped and ETT exchanged trigonal pyramidal CIS NCs (Δ-CIS). A well-known feature is the olefinic (-HC=CH-) stretching mode of the double bond in OLA at 3006 cm$^{-1}$.[27] This stretching mode is observable in the OLA stabilized sample and completely absent in all ETT stabilized NC samples (Figure 2 and Figure S7). Distinct asymmetric and symmetric -CH$_2$ stretching vibrations of OLA occur near 2925 and 2850 cm$^{-1}$ in OLA capped samples (Figure 2 and S8). After ETT exchange the -CH$_3$ vibration of the ethyl-group in ETT at 2950 cm$^{-1}$ emerges. The -CH$_2$ vibration at 2850 cm$^{-1}$ is still visible with lower intensity (only one -CH$_2$ group in ETT). OLA stabilized samples show broad N-H vibrations at 3250 cm$^{-1}$ and 1650 cm$^{-1}$ (Figure 2 and S8). These N-H vibrations are completely absent in ETT capped NC samples and confirm the assumption that ligand exchange of OLA by ETT was successful. ETT capped NCs show the weak stretching vibration of the tetrazole rings bound to the NC surface near 1020-1160 cm$^{-1}$.[28,29] As described by Voitekhovich et al. these vibrations tend to alter due to different binding possibilities of the tetrazole rings to the NC surface. Furthermore, in ETT capped NC samples the distinct vibration of the pure and unbound ligand at 3067 cm$^{-1}$ is absent. This might be due to reduced vibration possibility of ETT when bound to a NC surface.

*2.2.2. Thermogravimetric Analysis (TGA) and Electron Ionization Mass Spectrometry (EI-MS)*

As ETT is known to decompose at moderate temperatures we followed the thermolysis of the ligand off NC surfaces by Thermogravimetric Analysis (TGA). NC samples were heated at a rate of 20 °C/min up to 550 °C under nitrogen flow and between 550 °C to 900 °C under a nitrogen/oxygen environment to remove residual carbon contents as carbon dioxide. **Figure 3**a shows the thermolysis of pure ETT with a single mass loss (94 %) at 218 °C visible in the first derivative of the measurement and attributed to the decomposition of the ligand. At ~700 °C remainders dissociate as volatile species under oxidative conditions (see Figure S9 for first derivatives of ETT capped CIS and CIGS NCs). Figure 3b depicts that ETT capped CIS and CIGS NCs exhibit the same decomposition trend with a first mass loss attributed to ETT thermolysis between 218 – 255 °C. All ETT capped NCs undergo ligand thermolysis at lower temperatures than OLA capped NC samples (see Figure S10).

Electron Ionization Mass Spectrometry (EI-MS) elucidates the decomposition way of ETT. With respect to a big fragment ion at *m/z* (%) 59 (68) [HSCN$^+$] originating from thiocyanic acid we can confirm the ring fragmentation reaction for cadmium 1-ethyl-5-thiotetrazole proposed by Voitekhovich et al.[25] An azide based decomposition of ETT is suggested as the spectrum lacks nitrogen associated mass peaks but shows fragment ions at *m/z* (%) 87 (8) [HSCN$_3^+$] as azide and sulfur containing part of the tetrazole ring and protonated azide at *m/z* (%) 43 (15) [HN$_3^+$].[30,31] Using the combined spectroscopic, thermogravimetric and spectrometric data we conclude that ETT decomposes virtually completely under the described conditions leaving the surface of CIS and CIGS NCs organics-free with residual sulfur (EDS see Figure S11).

*2.2.3. Small Angle X-Ray Scattering (SAXS)*

To examine long-range ordering of ETT capped NCs in comparison with initial OLA capped samples Small Angle X-ray scattering (SAXS) measurements were performed. SAXS reveals that the interparticle distance of NCs decreases after ligand exchange from OLA to ETT and underpins the successful procedure. The center-to-center distance of elongated CIS NCs (size distribution: 5.6 nm ± 1.4 nm in length) decreases by 4.7 nm after ETT exchange (**Figure 4**a). This high value originates from the broad scattering profile and leads to certain inaccuracies in NC size distribution and in OLA length and spacing estimation. However, after ligand exchange ETT capped samples show significantly improved ordering visible in the narrowed peak width of the scattering curve. After heating the samples to 260 °C for 90 minutes the center-to-center distance of the ligand-free NC films further decreases by 0.6 nm. At the same time the peak width broadens due to aggregation and beginning sintering processes in bare NC films. The center-to-center distance of CIGS NCs (7.4 ± 1 nm) decreases by 0.9 nm after ligand exchange (Figure 4b). The peak width is significantly narrowed indicating small size distribution of the NCs and improved ordering. After annealing of the sample at 260 °C the center-to-center distance of the NCs further decreases by 0.7 nm and the peak width of the heated samples broadens like described for CIS NCS (see TEM heating experiments, **Figure 5**). SAXS patterns of trigonal pyramidal CIS NCs are excluded as the particles have a broad size distribution and lack distinct long-range ordering.

SAXS results show that elongated CIS and CIGS NCs both feature improved long range ordering after ligand exchange upon ETT accentuating the beneficial role of ETT as ligand for the chalcopyrite systems.

*2.2.4. TEM Heating Experiments*

To trace the exact processes during the NC thin film annealing step and the coalescence of the NCs we performed stepwise heating of ETT capped CIGS NCs inside the TEM. At every heating equilibration step EDS was performed (see Figure S11). The materials' composition and chalcopyrite crystal phase thereby are unaltered in the temperature range up to 400 °C also seen in the selected area diffraction of the sample (Figure S12). Figure 5 shows CIGS NCs at different temperatures. A reduced interparticle distance as determined by SAXS is also apparent in the TEM images at 250 °C, the temperature where ETT starts to decompose. At 400 °C the NCs have started sintering and form assemblies. During heating of the CIGS NCs residual sulfur from ETT exchange is completely removed within the heating range (see Figure S11).

Comprehensive analysis of the ligand exchange processes of CIS and CIGS NCs with ETT involving ATR-FTIR, TGA, SAXS and TEM heating experiments emphasize the successful functionalization of the materials. ATR-FTIR measurements and TGA underpin the complete ligand exchange. SAXS measurements reveal that after ligand exchange interparticle distances are reduced and improved long-range ordering of elongated CIS and CIGS is apparent. Heating experiments inside TEM show that after thermolysis of ETT sintering of virtually bare surfaced and organic ligand-free CIGS NCs is possible without altering the chalcopyrite crystal phase and stoichiometry of the material.

**2.3. Electrical Transport Measurements**

Electrical transport in closely assembled NC solids is highly dependent on NC surface functionalization. Long-chained organic capping agents like OLA suppress effective interparticle coupling and lead to high tunnelling barriers for electrons and thus insulating behavior of the materials.[7] Reduced interparticle distance improves particle coupling and hence charge transport in NC assemblies. To monitor the effect of reduced interparticle spacing upon ETT exchange, we investigate the electrical transport behavior of the chalcopyrite NCs before and after ETT thermolysis. The current-voltage characteristics of trigonal pyramidal and elongated CIS thin films as well as CIGS thin films are measured in the dark and under illumination. As ETT undergoes thermolysis at different temperatures for the three systems (see Figure 3b and Figure S9) all NC thin film samples were heated to 260 °C to assure complete ETT decomposition and comparable conditions but prevent sintering of the NCs as observed at 400 °C. In a typical experiment 20 μL (2 x 10 μL) of a colloidal CIS or CIGS NC solution were spin-coated (20 sec at 500 rpm, subsequently 20 sec at 2000 rpm) onto a highly n-doped silicon substrate with a 300 nm insulating layer of silicon oxide and pre-deposited gold electrodes. Directly after spin-coating the substrates were transferred to a probe station connected with a parameter analyzer and electrical measurements were carried out under vacuum conditions. The NC thin films are characterized in the dark and under illumination with a Schott ACE 150 Watt halogen AC powered light source with fiber optics. The bias voltage $V_{DS}$ is swept between -5 V and +5 V. After measuring the current-voltage curves of ETT capped NCs the samples were transferred to an oven and heated to 260 °C for 90 minutes under oil pump vacuum. The annealed samples were kept under argon as they were prone to oxidation and transferred back to the probe station. The previously characterized devices are measured again in the dark and under illumination. The conductivity ($\sigma$) of the samples is then calculated according to

$$\sigma = \frac{I_{ds}}{V_{ds}} \cdot \frac{L}{h \cdot W} \tag{1}$$

with $V_{ds}$ = +200 mV, channel length $L$ = 300 nm for trigonal pyramidal CIS, 350 nm for elongated CIS and CIGS, width $W$ = 14.5 µm and height $h$ of the film ~20 nm for trigonal pyramidal CIS and ~40 nm for elongated CIS and CIGS films. NC thin films which were exemplarily heated to 400 °C to enable sintering of the NCs were excluded from current-voltage measurements as they featured intercalation of the NCs into the electrode setup. Different electrode settings are under investigation to address this problem. **Figure 6**a shows the current-voltage characteristics of a trigonal pyramidal NC thin film before and after ETT thermolysis on the logarithmic scale (Figure S13-S15 for SEM images of the NC thin films).

**Table 1** shows the conductivities of ETT capped CIS and CIGS NCs. Prior to ligand thermolysis trigonal pyramidal CIS NC exhibit a conductivity of $1.4 \times 10^{-9}$ S/cm in the dark. These values are comparable to conductivities exemplarily measured for trigonal pyramidal CIS NCs capped with the initial ligand OLA under same conditions. After ETT thermolysis however the NC film shows a significant four orders of magnitude enhanced electrical transport with a conductivity of $1.4 \times 10^{-5}$ S/cm under illumination. Elongated CIS NC films were characterized likewise (see Table 1 and Figure 6b) exhibiting conductivities of $3.7 \times 10^{-9}$ S/cm in the dark prior to ligand thermolysis. The hysteresis in the current-voltage curves of ETT capped samples prior to heating (Figure 6a-c) originates from charging of the NC film when a bias voltage of -5V to +5V is applied. However, after annealing this effect is minimized once more underpinning the successful ligand-free state of the NC films. Electrical transport in elongated ligand-free CIS NCs is enhanced by three orders of magnitude to conductivities of $6.8 \times 10^{-6}$ S/cm under illumination. In comparison with trigonal pyramidal NCs it is apparent that the NC size plays a crucial role in electrical transport of the CIS materials. Elongated CIS NCs with an average size of 9 nm show 1.5 orders of magnitude reduced conductivity compared to 19 nm sized trigonal pyramidal CIS NCs. Two factors are supposable for this behavior:

1) The higher surface-to-volume ratio in smaller and organic ligand-free NCs leads to more surface originating trapping states known to reduce electrical transport.[32]

2) Charge carriers in smaller NCs have to overcome more grain boundaries to cross the channel length and will likely more often recombine.

Figure 6c shows electrical transport of spherical CIGS NC films exhibiting conductivities of $1.0 \times 10^{-9}$ S/cm in the dark prior to ligand thermolysis. After annealing the same devices show two orders of magnitude enhanced conductivity of $1.9 \times 10^{-7}$ S/cm under illumination. Here again the higher number of trap states and grain boundaries in smaller NCs may lead to a lower conductivity. All NC types show photoresponse after ligand thermolysis whereas CIGS NCs exhibit the strongest effect. The current measured at +5 V increases from $5 \times 10^{-10}$ to $1 \times 10^{-9}$ Ampere, a 2 fold gain, when ligand-free NCs are illuminated by diffuse microscope light (see Figure S16C).

## 3. Conclusion

We have demonstrated the successful functionalization of CIS and CIGS NCs with ETT as thermally degradable ligand preserving the NC's colloidal stability prior to ligand thermolysis and obtaining organics-free individual NCs after the annealing step. By this straight forward, facile and save method for the first time we investigated CIS and CIGS NC solids' significantly enhanced electrical transport in a virtually ligand-free state. These NCs can be sintered to form close crystalline assemblies without altering the chalcopyrite crystal structure and composition of the material. The combination of synthesis, detailed characterization and application of the chalcopyrites yields CIS and CIGS NC materials that hold high potential for the utilization in photovoltaic devices.

## 4. Experimental

*Chemicals*: All Chemicals were used as received without further purification. Copper(I) acetate (Cu(OAc), STREM, 99%), copper(I) chloride (CuCl, Sigma Aldrich, 99.995%), indium(III) chloride (InCl$_3$, Sigma Aldrich, 99.99%), gallium (III) iodide (GaI$_3$, Sigma Aldrich, 99.999%), selenourea (Sigma Aldrich, 98%) and oleylamine (OLA, Acros Organics, 80-90%) were stored in a glovebox in a nitrogen environment. Trioctylphosphine (TOP, 90%) was purchased from Sigma Aldrich and stored under ambient conditions.

*Methods:* Unless stated otherwise all synthetic steps were carried out using standard Schlenk-line techniques under nitrogen as inert gas or inside a nitrogen filled glovebox.

*Synthesis of trigonal pyramidal CuInSe$_2$:* Trigonal pyramidal CuInSe$_2$ NCs were synthesized according to a method of Koo et al.[26] In a typical synthesis of 19 nm NCs 0.50 mmol (50 mg) copper(I) chloride were mixed with 0.500 mmol (111 mg) indium(III) chloride inside a glovebox in a 25 mL threenecked flask equipped with a septum and thermocouple. 10 mL OLA were added to the components and the solution marked A. The flask was connected to a condenser under nitrogen flow outside of the glovebox. In a separate 5 mL threenecked flask likewise equipped with septum, thermocouple and condenser, 1.00 mmol (123 mg) selenourea was mixed with 1 mL OLA and marked solution B. Solution A was kept stirring under oil pump vacuum at 60 °C for one hour and solution B was stirred under nitrogen flow. After one hour solution A was set under nitrogen, heated to 130 °C for 10 minutes and subsequently cooled to 100 °C. Solution B meanwhile was heated to 210 °C to dissolve the reactant and set free reactive selenium species for reaction. Subsequently it was cooled to 110 °C, taken out by syringe and rapidly injected into solution A. The reaction mixture was then heated to 240 °C with a rate of 15 °C/min. After one hour of reaction time the suspension was cooled to room temperature. 30 mL of ethanol were added to the crude product to precipitate the NCs. The mixture was centrifuged at 4000 rpm for 4 minutes. The colorless supernatant was discarded and the NCs redissolved in 5 mL chloroform. After centrifugation of this solution at 4000 rpm for 5 minutes the NC product in the deep black supernatant was separated from sediment which contained poorly capped NCs. The reaction typically yielded ~50 % purified NCs. Characterization and ligand exchange was conducted with the purified NCs from the supernatant (Figure S1 and S2).

*Synthesis of elongated CuInSe$_2$:* Elongated CuInSe$_2$ NCs were synthesized based on a method by Panthani et al.[4] A typical synthesis included 0.50 mmol (61 mg) copper(I) acetate, 0.500 mmol (111 mg) indium(III) chloride and 1.00 mmol (123 mg) selenourea mixed in a 25 mL threenecked flask and combined with 10 mL OLA. The reaction solution was kept under oil pump vacuum at 60 °C for one hour, then set under nitrogen and stirred for another hour at 110 °C. Subsequently the mixture was heated to 240 °C and kept at this temperature for one hour. Afterwards the solution was cooled to room temperature. For purification and better stabilization of the NC product 4 mL of a TOP/toluene mixture (TOP/Tol 3:1) was added to the crude product. The solution was vortexed and centrifuged at 4500 rpm for 5 minutes to remove poorly capped NCs. The supernatant was separated from the sediment and 30 mL 2-propanol was added for precipitation. The mixture was centrifuged at 10000 rpm for 15 minutes. The NC product was redissolved in 5 mL chloroform and filtrated with a 0.2 μm hydrophobic syringe filter. 15 mL of a methanol/2-propanol solution (4:1) were added and the mixture centrifuged at 10000 rpm for 5 minutes. The reaction typically yielded ~25 % (Figure S3 and S4). NCs were redissolved in chloroform and used for characterization and ligand exchange.

*Synthesis of CuIn$_{1-x}$Ga$_x$Se$_2$:* CuIn$_{1-x}$Ga$_x$Se$_2$ NCs were synthesized based on a method by Panthani et al.[4] In a typical one-pot synthesis 0.10 mmol (12 mg) copper(I) acetate, 0.10 mmol (22 mg) indium(III) chloride, 0.10 mmol (45 mg) gallium(III) iodide and 0.20 mmol (25 mg) selenourea were mixed in a 25 mL threenecked flask and 10 mL OLA were added. The reaction mixture was stirred at 60 °C under oil pump vacuum, then set under nitrogen and stirred at 110 °C for an additional hour and subsequently heated to 240 °C for one hour for reaction. Afterwards the solution was allowed to cool to room temperature. For purification and better stabilization of the NC product 4 mL of TOP/Tol (3:1) were added to the crude product. The solution was vortexed and centrifuged at 5600 rpm for 5 minutes to remove poorly capped NCs. The supernatant was separated and 25 mL 2-propanol were added for precipitation. The mixture was centrifuged at 10000 rpm for 15 minutes. The NCs were redissolved in 2 mL chloroform and filtrated with a 0.2 μm hydrophobic syringe filter. The reaction typically yielded ~60 % (Figure S5 and S6).

*1-ethyl-5-thiotetrazole:* 1-ethyl-5-thiotetrazole was synthesized by cycloaddition of sodium azide to ethyl isothiocyanate according to well established literature procedures[33,34].

*Transmission Electron Microscopy:* TEM imaging was conducted with a JEOL JEM 2200FS (UHR) with CESCOR and CETCOR corrector at an acceleration voltage of 200 kV or a JEOL JEM 1011 microscope at 100 kV equipped with a CCD camera.

*Energy Dispersive X-Ray Spectrometry:* Quantitative elemental analysis of NCs was obtained using a Si(Li) JEOL JED-2300 energy dispersive X-ray detector.

*Scanning Election Microscopy:* SEM images were obtained on a LEO1550 scanning electron microscope with a spatial resolution of ∼1 nm.

*X-Ray Diffraction:* Powder XRDs were recorded using a Philipps X'Pert-diffractometer with Bragg-Brentano-geometry on applying copper-Kα radiation.

*Attenuated Total Reflectance Fourier Transform Infrared Spectroscopy (ATR-FTIR):* IR spectra were recorded with a *Bruker Equinox 55* spectrometer applying Attenuated Total Reflectance Fourier Transform Infrared Spectroscopy (ATR-FTIR) technique. The dried NC sample was placed on the crystal of the ATR unit and the spectra recorded in the range of 600-4000 $cm^{-1}$.

*Small Angle X-Ray Scattering:* SAXS experiments were performed at a rotating anode device consisting of a rotating copper (Cu) anode (Seifert, Ahrensburg), crossed Goebel mirrors, and an image plate detector (Fuji) at a sample–detector distance of 1.0 m.

*Thermogravimetric Analysis:* TGA studies were carried out using a *Netzsch TG 209 C* in the temperature range of 25 – 550 °C with a heating rate of 20 °C/min under nitrogen atmosphere. Subsequently samples were heated from 550 – 900 °C under oxygen/nitrogen atmosphere at a rate of 20 °C/min.

*Characterization of Electrical Transport and Conductivity Measurements:* The room temperature electrical measurements were performed on a 4200-SCS semiconductor characterization system from Keithley Instruments inside the VFTTP4 probestation by Lake Shore Cryotronics.

**Acknowledgements**

We gratefully acknowledge Jan Michels for the possibility to use the Atomic Layer Deposition system of the Bachmann group for heating of the NC coated $Si/SiO_2$ substrates. Carsten Ott from the Center for Applied Nanotechnology (CAN) Company in Hamburg, Germany is gratefully acknowledged for synthesizing and providing the 1-ethyl-5-thiotetrazole (ETT) ligand.

# References

[1] F. Hergert, S. Jost, R. Hock, M. Purwins, *J. Solid State Chem.* **2006**, *179*, 2394.
[2] W. Shockley, H. J. Queisser, *J. Appl. Phys.* **1961**, *32*, 510.
[3] B. J. Stanbery, *Crit. Rev. Solid State Mater. Sci.* **2002**, *27*, 73.
[4] M. G. Panthani, V. Akhavan, B. Goodfellow, J. P. Schmidtke, L. Dunn, A. Dodabalapur, P. F. Barbara, B. A. Korgel, *J. Am. Chem. Soc.* **2008**, *130*, 16770.
[5] J. Tang, S. Hinds, S. O. Kelley, E. H. Sargent, *Chem. Mater.* **2008**, *20*, 6906.
[6] C. B. Murray, C. R. Kagan, M. G. Bawendi, *Ann. Rev. Mater. Sci.* **2000**, *30*, 545.
[7] D. V. Talapin, J.-S. Lee, M. V. Kovalenko, E. V. Shevchenko, *Chem. Rev.* **2010**, *110*, 389.
[8] V. I. Klimov, A. A. Mikhailovsky, S. Xu, A. Malko, J. A. Hollingsworth, C. A. Leatherdale, H.-J. Eisler, M. G. Bawendi, *Science* **2000**, *290*, 314.
[9] D. V. Talapin, C. B. Murray, *Science* **2005**, *310*, 86.
[10] V. A. Akhavan, M. G. Panthani, B. W. Goodfellow, D. K. Reid, B. A. Korgel, *Opt. Express* **2010**, *18*, A411.
[11] Q. Guo, S. J. Kim, M. Kar, W. N. Shafarman, R. W. Birkmire, E. A. Stach, R. Agrawal, H. W. Hillhouse, *Nano Lett.* **2008**, *8*, 2982.
[12] I. Gur, N. A. Fromer, M. L. Geier, A. P. Alivisatos, *Science* **2005**, *310*, 462.
[13] K. W. Johnston, A. G. Pattantyus-Abraham, J. P. Clifford, S. H. Myrskog, D. D. MacNeil, L. Levina, E. H. Sargent, *Appl. Phys. Lett.* **2008**, *92*, 151115.
[14] F. Shen, W. Que, P. Zhong, J. Zhang, X. Yin, *Coll. Surf. A: Physicochem. Eng. Asp.* **2011**, *392*, 1.
[15] P. Geiregat, Y. Justo, S. Abe, S. Flamee, Z. Hens, *ACS Nano* **2013**, *7*, 987.
[16] M. C. Beard, J. M. Luther, O. E. Semonin, A. J. Nozik, *Acc. Chem. Res.* **2012**, DOI:10.1021/ar3001958.
[17] O. E. Semonin, J. M. Luther, S. Choi, H.-Y. Chen, J. Gao, A. J. Nozik, M. C. Beard, *Science* **2011**, *334*, 1530.
[18] A. de Kergommeaux, A. Fiore, J. Faure-Vincent, F. Chandezon, A. Pron, R. de Bettignies, P. Reiss, *Mater. Chem. Phys.* **2012**, *136*, 877.
[19] M. V. Kovalenko, M. Scheele, D. V. Talapin, *Science* **2009**, *324*, 1417.
[20] J. M. Luther, M. Law, Q. Song, C. L. Perkins, M. C. Beard, A. J. Nozik, *ACS Nano* **2008**, *2*, 271.
[21] D. A. R. Barkhouse, A. G. Pattantyus-Abraham, L. Levina, E. H. Sargent, *ACS Nano* **2008**, *2*, 2356.
[22] A. T. Fafarman, W. Koh, B. T. Diroll, D. K. Kim, D.-K. Ko, S. J. Oh, X. Ye, V. Doan-Nguyen, M. R. Crump, D. C. Reifsnyder, C. B. Murray, C. R. Kagan, *J. Am. Chem. Soc.* **2011**, *133*, 15753.
[23] E. L. Rosen, R. Buonsanti, A. Llordes, A. M. Sawvel, D. J. Milliron, B. A. Helms, *Angew. Chem. Int. Ed.* **2012**, *51*, 684.
[24] H. Zhang, B. Hu, L. Sun, R. Hovden, F. W. Wise, D. A. Muller, R. D. Robinson, *Nano Lett.* **2011**, *11*, 5356.
[25] S. V. Voitekhovich, D. V. Talapin, C. Klinke, A. Kornowski, H. Weller, *Chem. Mater.* **2008**, *20*, 4545.
[26] B. Koo, R. N. Patel, B. A. Korgel, *J. Am. Chem. Soc.* **2009**, *131*, 3134.
[27] N. Shukla, E. B. Svedberg, J. Ell, *Crit. Rev. Solid State Mater. Sci.* **2007**, *301*, 113.
[28] A. Gómez-Zavaglia, I. D. Reva, L. Frija, M. L. Cristiano, R. Fausto, *Journal of Molecular Structure* **2006**, *786*, 182.
[29] B. Sägmüller, P. Freunscht, S. Schneider, *J. Mol. Struct.* **1999**, *482-483*, 231.
[30] K. W. Paul, M. M. Hurley, K. K. Irikura, *J. Phys. Chem. A* **2009**, *113*, 2483.
[31] N. Piekiel, M. R. Zachariah, *J. Phys. Chem. A* **2012**, *116*, 1519.
[32] P. Nagpal, V. I. Klimov, *Nat Commun* **2011**, *2*, 486.

[33]  P. J. Kocienski, A. Bell, P. R. Blakemore, *Synlett* **2000**, *2000*, 365.
[34]  H. R. Meier, H. Heimgartner, *Methoden der Organischen Chemie (Houben-Weyl)*; Schumann, E., Ed.; George Thieme, Stuttgart, Germany, 1994; Vol. E8d, pp. 664–795.

**Figures**

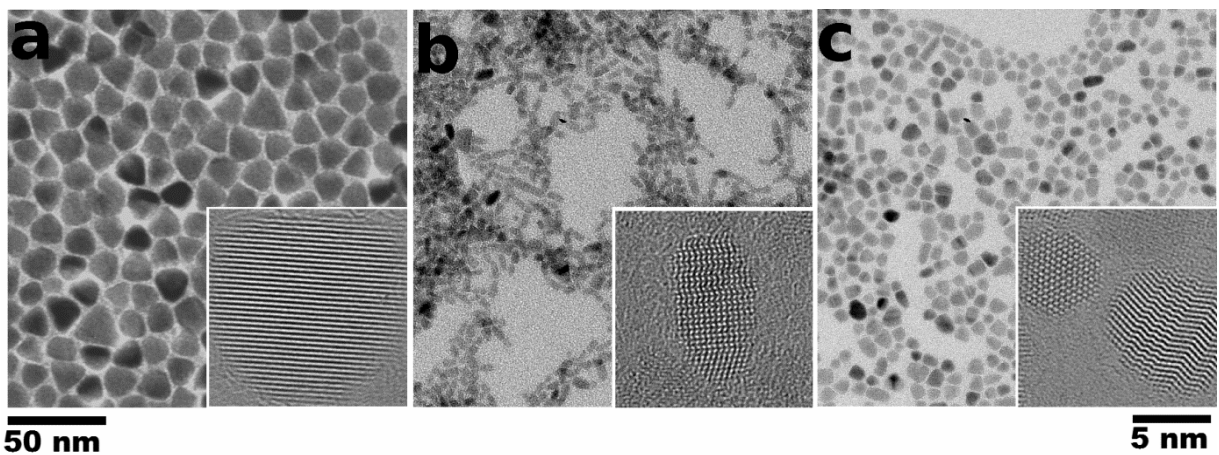

**Figure 1.** TEM images of ETT stabilized NCs, insets with single NCs (a) trigonal pyramidal CIS, (b) elongated CIS and (c) CIGS.

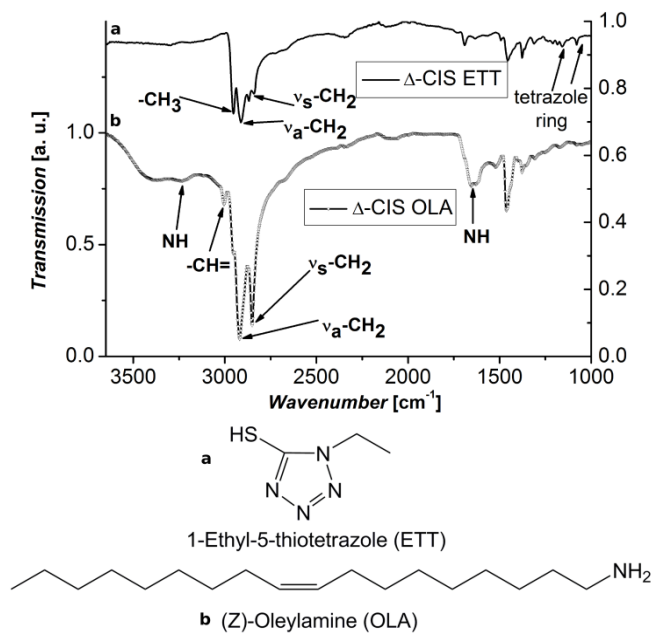

**Figure 2.** IR spectra of OLA capped and ETT exchanged NCs exemplarily shown for trigonal pyramidal NCs (Δ-CIS).

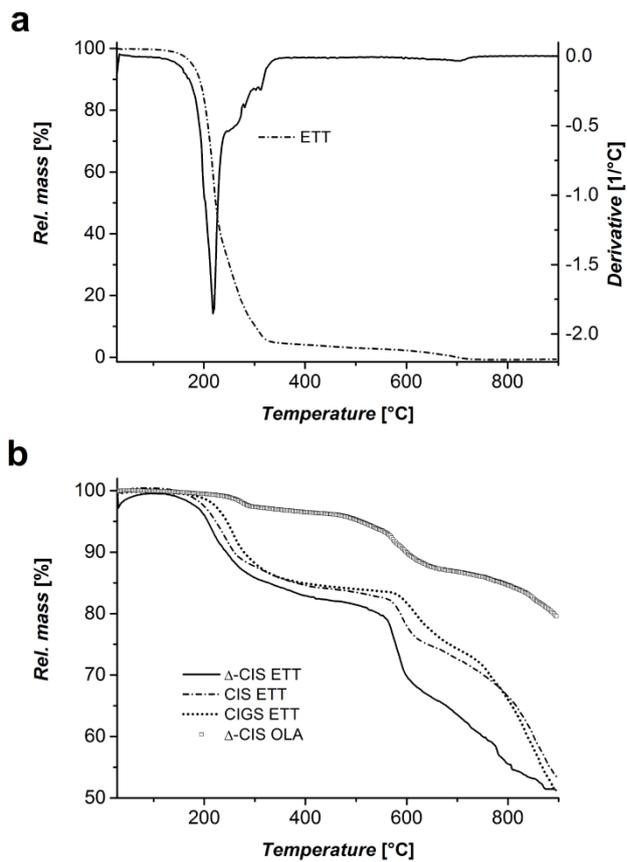

**Figure 3.** TGA of (a) pure ETT with thermolysis temperature of 218 °C, (b) ETT capped CIS and CIGS NCs and OLA capped trigonal pyramidal NC sample for comparison, all ETT capped NCs show a distinct first mass loss attributed to ETT thermolysis in the range of 218 – 255 °C.

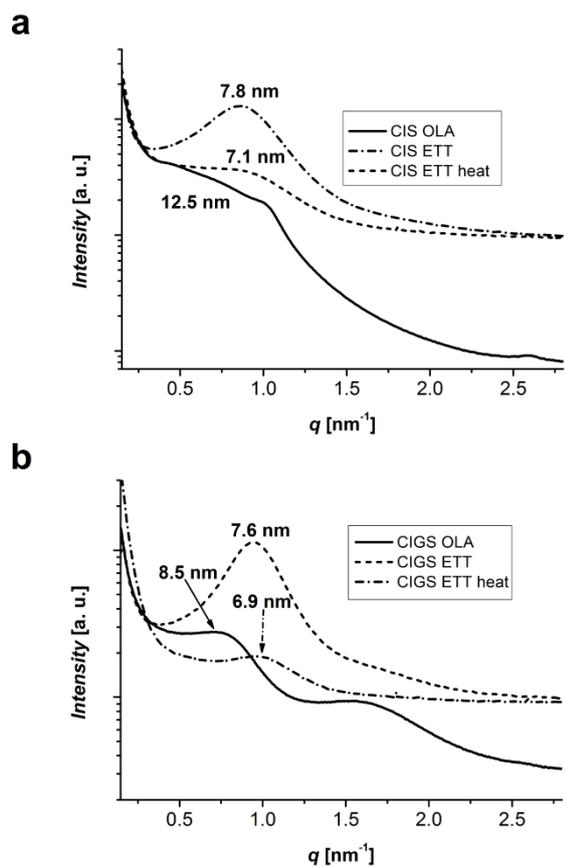

**Figure 4.** SAXS of (a) elongated CIS NCs, showing reduced center-to-center distance and improved ordering of NCs after ETT exchange (b) CIGS NCs prior to and after ligand exchange, center-to-center distance decreases from OLA capped to ETT exchanged to heated and organic ligand-free, ordering improves after ligand exchange and decreases after heating of the sample.

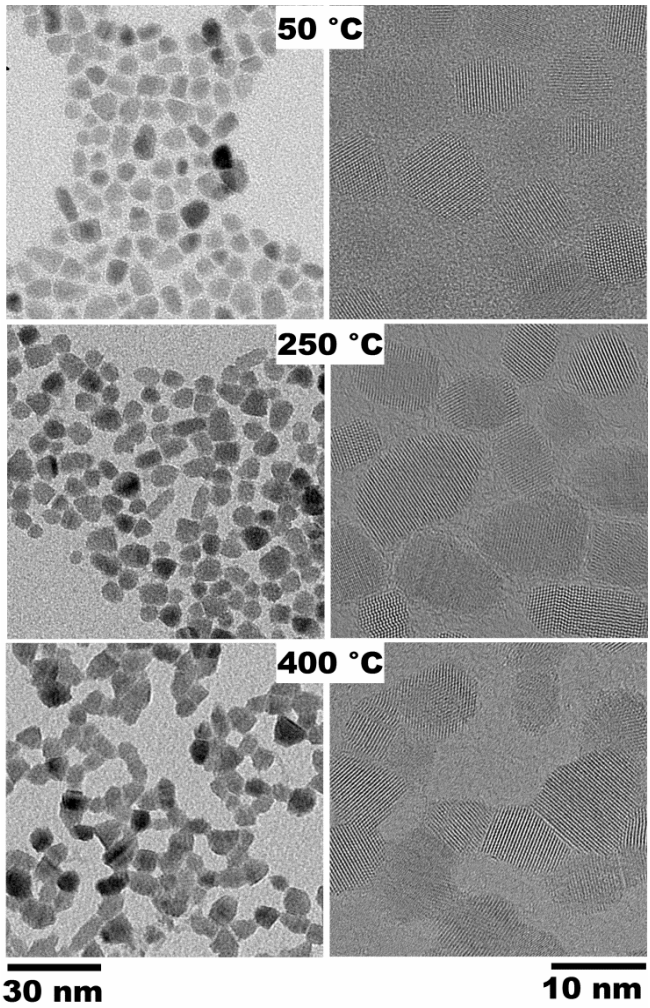

**Figure 5.** Heating experiment of CIGS NCs inside TEM at 50 °C, 250 °C with reduced interparticle spacing, at 400 °C NCs start sintering.

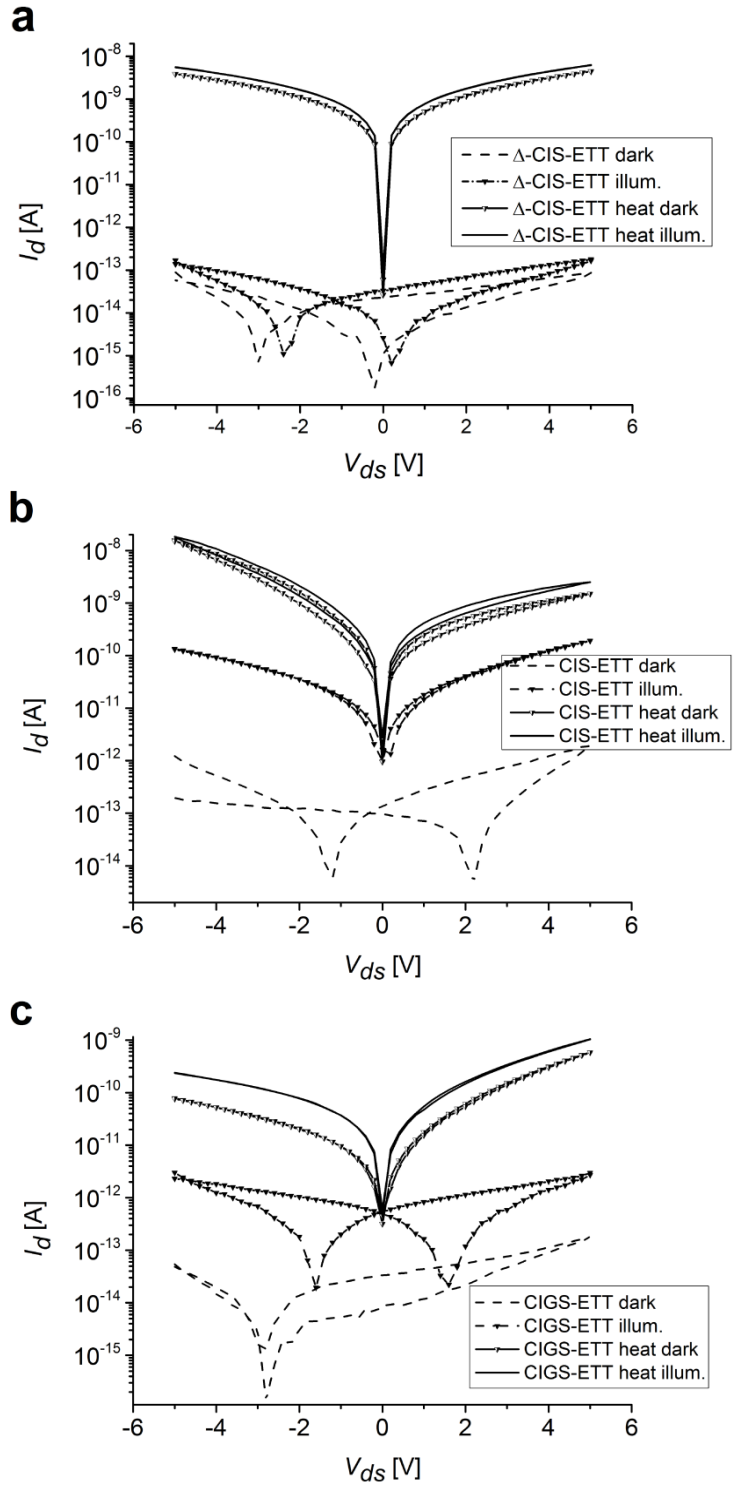

**Figure 6.** Current-voltage curves at room temperature: (a) trigonal pyramidal CIS NCs (Δ-CIS), conductivity increase by four orders of magnitude after thermolysis of ETT, (b) elongated CIS NCs and three orders of magnitude enhancement in conductivity after ETT thermolysis under illumination compared to conductivity in the dark prior to ETT thermolysis, (c) CIGS NCs, conductivity improvement by two orders of magnitude in total.

**Table 1.** Conductivity of CIS and CIGS NCs

|  | Trigonal pyramidal CIS $\sigma$ [S/cm] | Elongated CIS $\sigma$ [S/cm] | CIGS $\sigma$ [S/cm] |
|---|---|---|---|
| ETT dark | $1.4 \times 10^{-9}$ | $3.7 \times 10^{-9}$ | $1.0 \times 10^{-9}$ |
| ETT illuminated | $4.0 \times 10^{-9}$ | $6.8 \times 10^{-8}$ | $1.3 \times 10^{-8}$ |
| ETT heat dark | $7.5 \times 10^{-6}$ | $8.2 \times 10^{-7}$ | $6.0 \times 10^{-8}$ |
| ETT heat illuminated | $1.4 \times 10^{-5}$ | $6.8 \times 10^{-6}$ | $1.9 \times 10^{-7}$ |

Supporting Information

**Virtually Bare Nanocrystal Surfaces – Significantly Enhanced Electrical Transport in CuInSe$_2$ and CuIn$_{1-x}$Ga$_x$Se$_2$ Thin Films upon Ligand Exchange with Thermally Degradable 1-Ethyl-5-thiotetrazole**

*Jannika Lauth, Jakob Marbach, Andreas Meyer, Sedat Dogan,
Christian Klinke, Andreas Kornowski, and Horst Weller\**

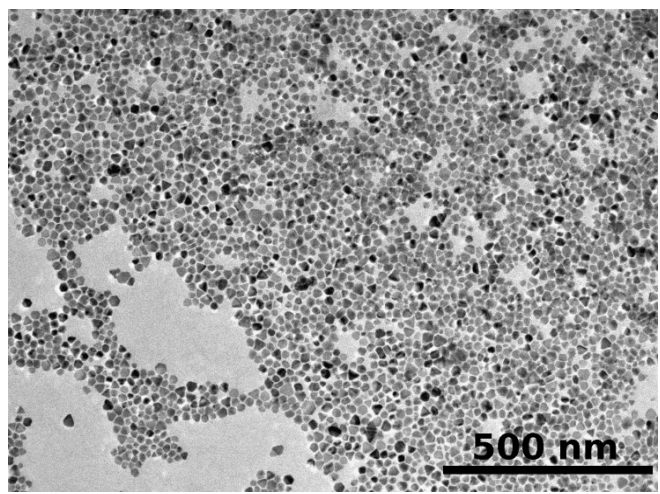

**Figure S1.** Trigonal pyramidal CIS NCs stabilized with OLA, prior to ligand exchange with ETT.

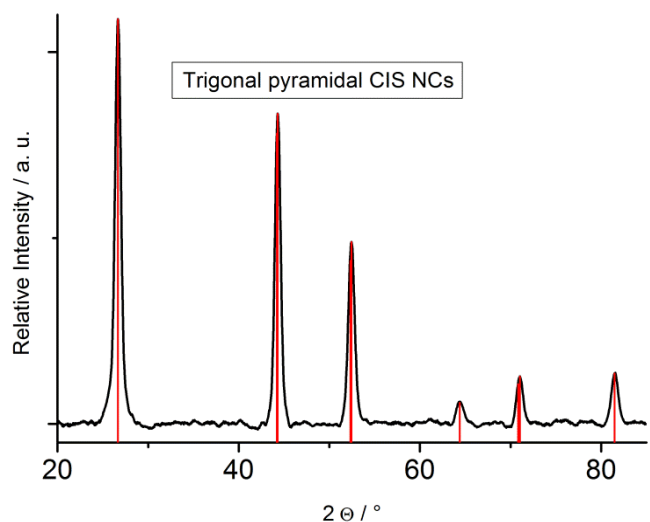

**Figure S2.** Powder XRD of trigonal pyramidal CIS NCs, solid lines indicating corresponding CuInSe$_2$ bulk reflexes (PDF card 00-040-1487).

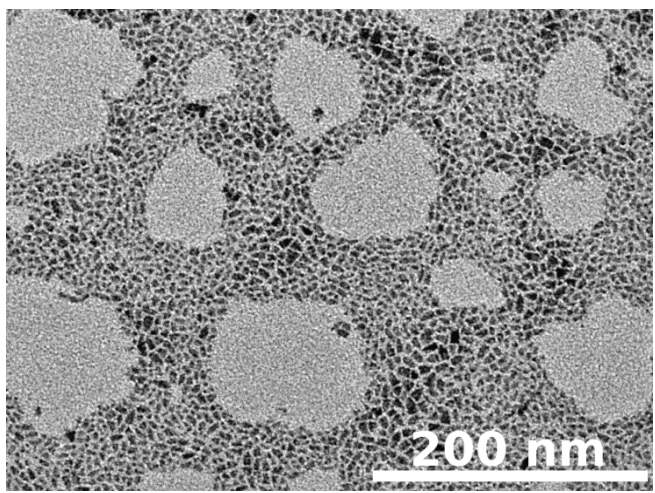

**Figure S3.** Elongated CIS NCs stabilized with OLA, prior to ligand exchange with ETT.

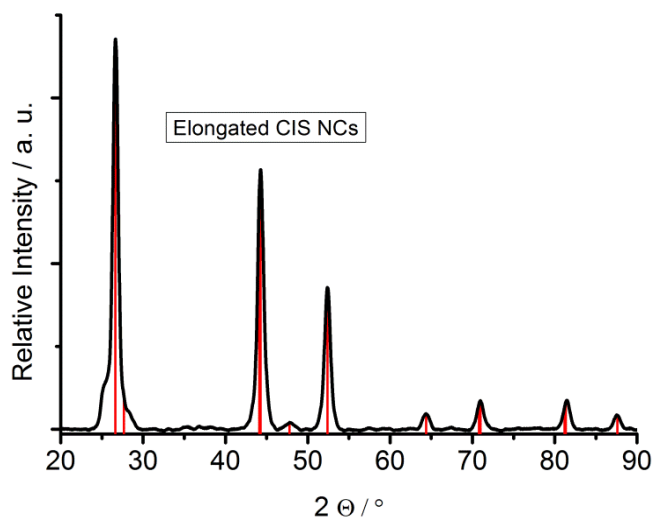

**Figure S4.** Powder XRD of elongated $CuInSe_2$, solid lines indicating bulk $CuInSe_2$ reflexes (PDF card 00-040-1487).

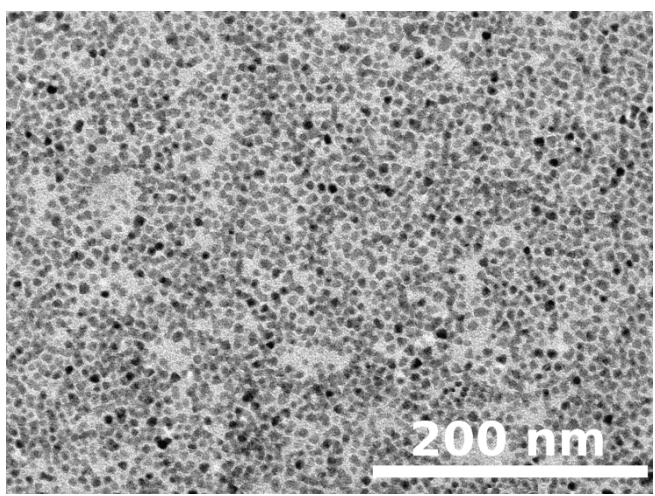

**Figure S5.** Spherical CIS NCs stabilized with oleylamine, prior to ligand exchange with ETT.

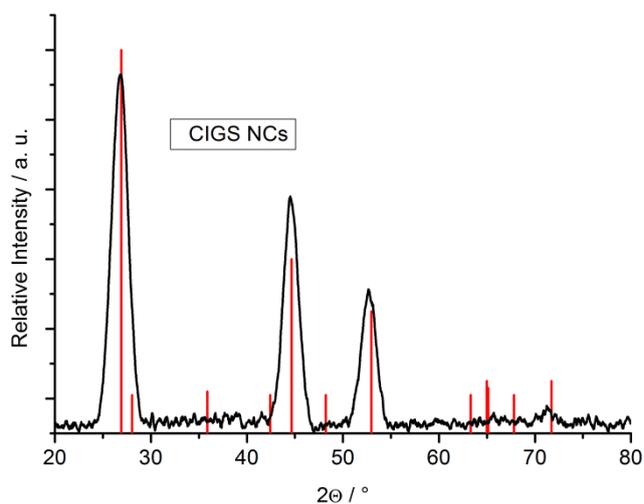

**Figure S6.** Powder XRD of spherical CuIn$_{1-x}$Ga$_x$Se$_2$, solid lines indicating bulk CuIn$_{0.7}$Ga$_{0.3}$Se$_2$ reflexes (PDF card 00-035-1102), closest match for the obtained NC lattice.

**Table S1. ETT Ligand Exchange Conditions for CIS and CIGS NCs.**

|  | Estimated NC concentration in CHCl$_3$ [mmol/mL] | ETT 2.8 M in CHCl$_3$ [mL] | Molar Excess ETT vs. NCs | Ligand Exchange Time [h] |
| --- | --- | --- | --- | --- |
| Trigonal pyramidal CIS | 0.025 | 2 | ~220 | 20 |
| Elongated CIS | 0.013 | 2 | ~450 | 20 |
| CIGS | 0.006 | 2 | ~930 | 30 |

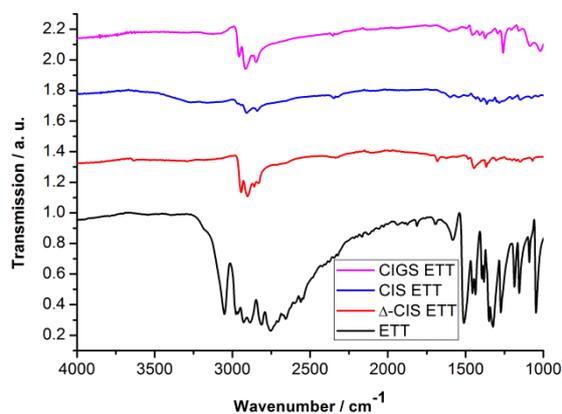

**Figure S7.** FT-IR spectra of pure ETT and all three ligand exchanged NC types.

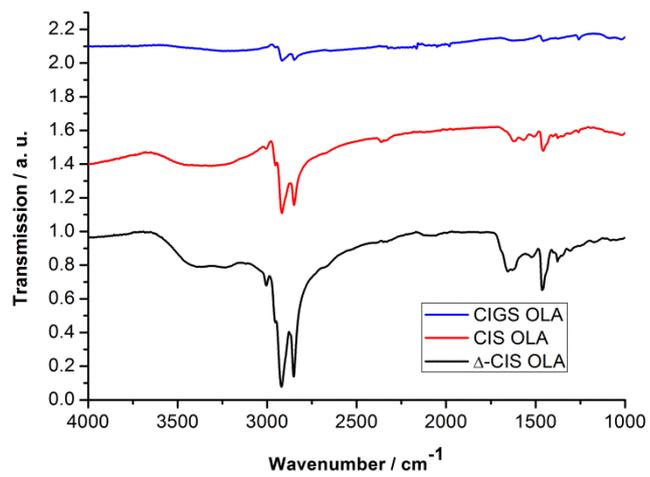

**Figure S8.** FT-IR spectra of OLA stabilized chalcopyrite NC samples.

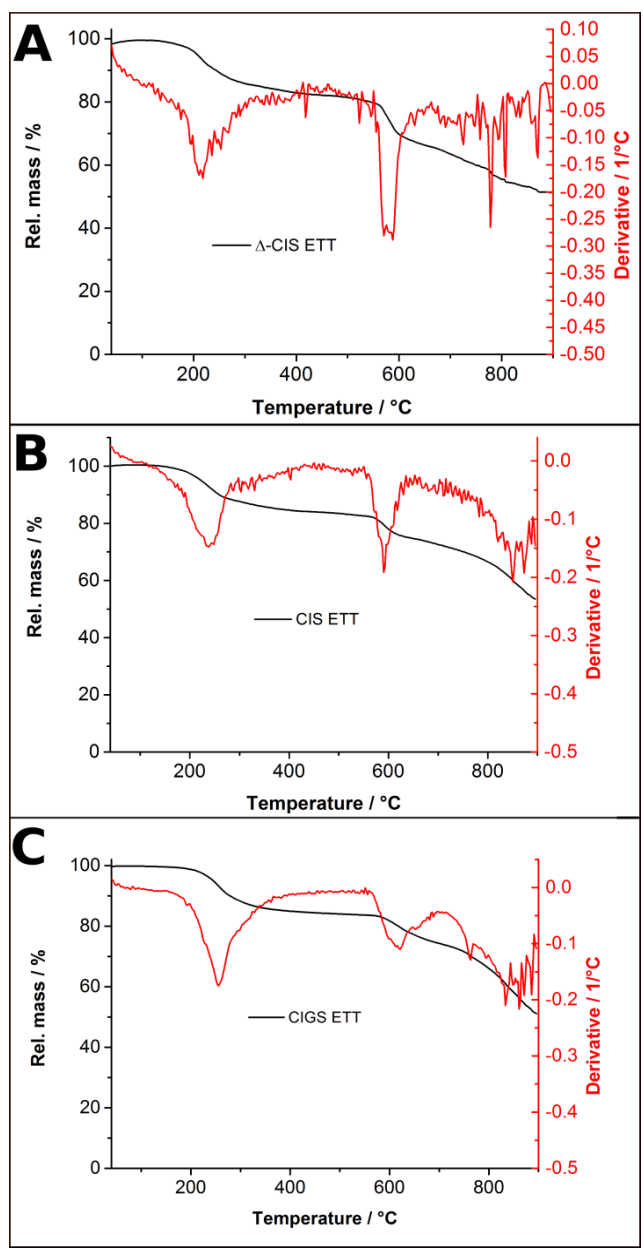

**Figure S9.** TGA of ETT capped (A) trigonal pyramidal CIS NCs (Δ-CIS) with ETT thermolysis at 218 °C, (B) elongated CIS NCs with thermolysis of the ligand at 237 °C and (C) CIGS NCs with thermolysis of the ligand at 255 °C.

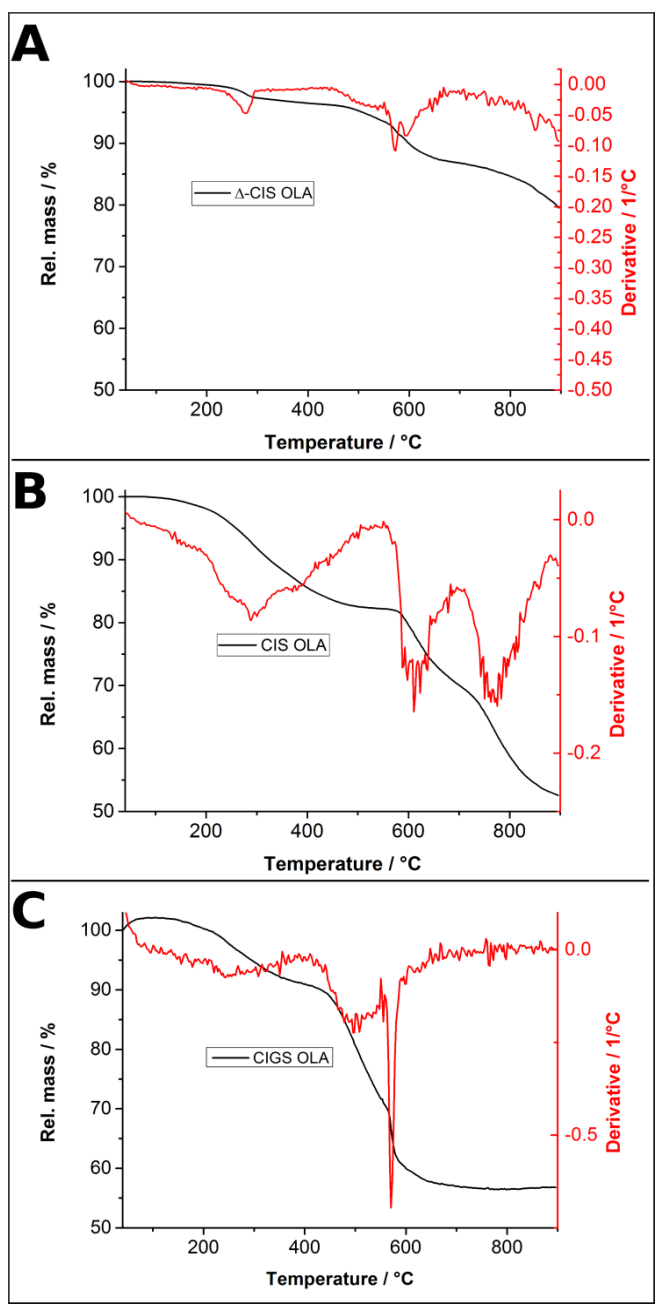

**Figure S10.** OLA capped chalcopyrite NCs, mass losses visible in the first derivative of the measurement curve (A) for trigonal pyramidal CIS NCs at ~278 °C, (B) for elongated CIS NCs at ~288 °C, (C) for CIGS NCs at around ~265-270 °C.

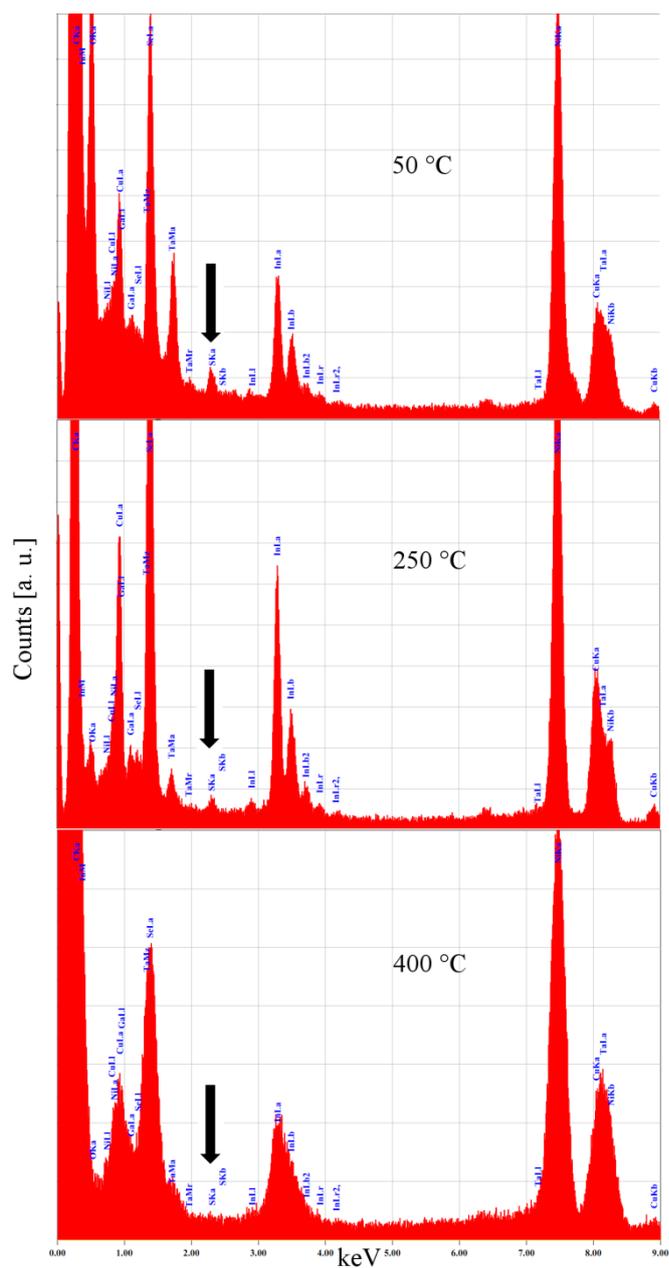

**Figure S11.** EDS showing unaltered CuIn$_{1-x}$Ga$_x$Se$_2$ composition with increasing temperature and reduced sulfur amount at higher temperatures.

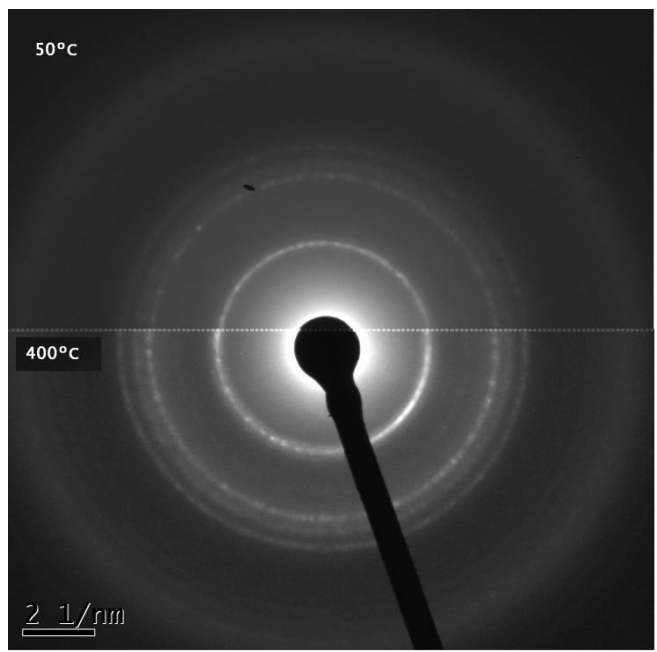

**Figure S12.** Selected Area Electron Diffraction (SAED) of inside TEM heated CIGS NCs, materials' composition and crystal phase stay unaltered between 50 °C and 400 °C.

To calculate the conductivity of a NC film we determined the film thickness of the different NC films by cross sectional SEM measurements. We tilted the stage to 70 ° and as we know that the gold electrodes on top of the silicon substrates have a height of 22 nm, the height of the assembly is ~20 nm. We therefore obtain for a typical monolayer of trigonal pyramidal NCs a height of 20 nm (Figure S13). Elongated CIS and CIGS thin films are estimated to have a height of ~40 nm (Figure S14 and S15).

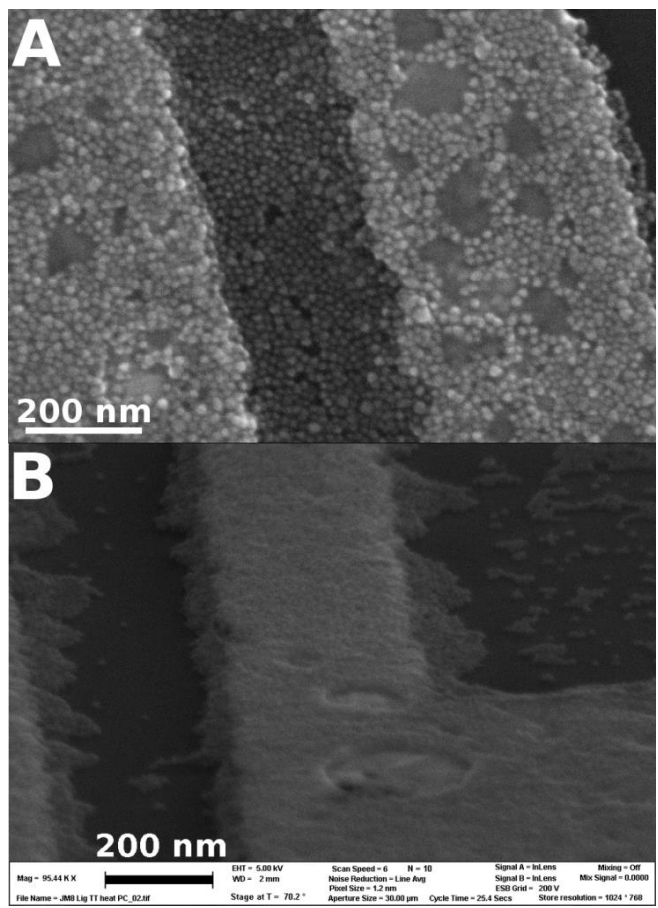

**Figure S13.** (A) Monolayer of trigonal pyramidal CIS NCs between two gold electrodes, top view, (B) 70 ° tilted view of a different electrode assembly, estimation of film thickness ~20 nm.

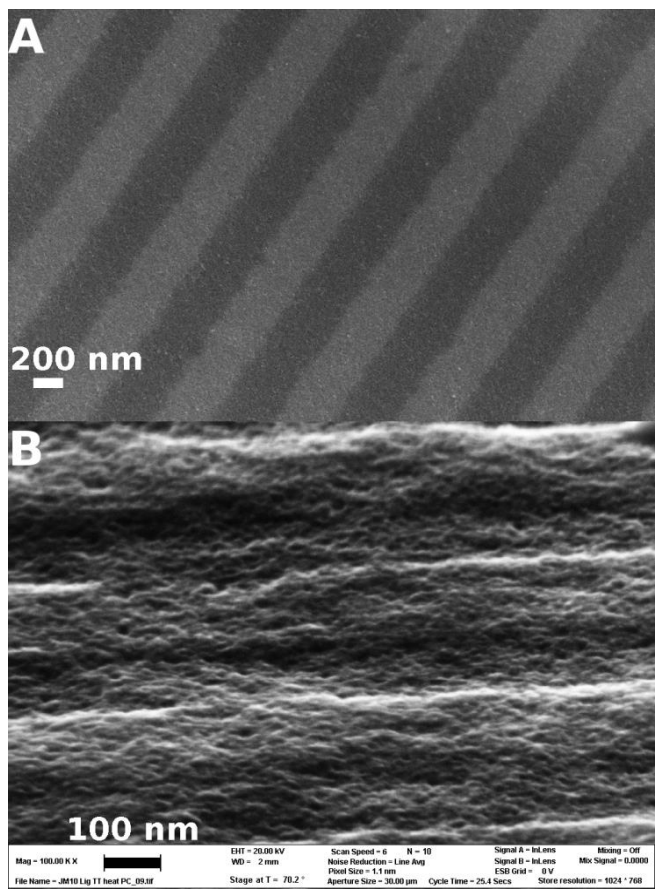

**Figure S14.** (A) Densely packed film of elongated CIS NCs, top view, (B) 70 ° tilted view on the same film.

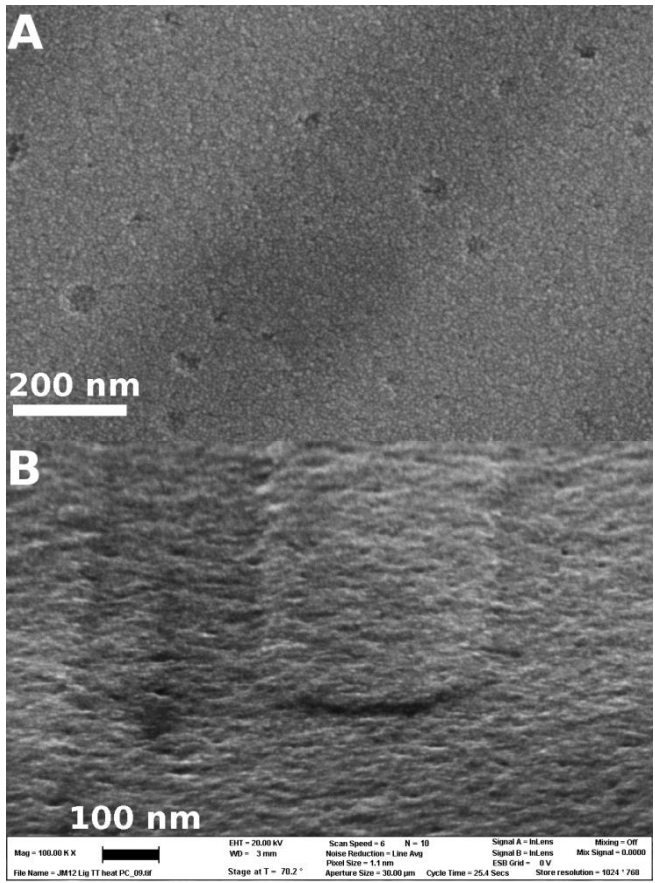

**Figure S15.** (A) densely packed film of CIGS NCs between two gold electrodes, (B) 70 ° tilted view on CIGS NC film.

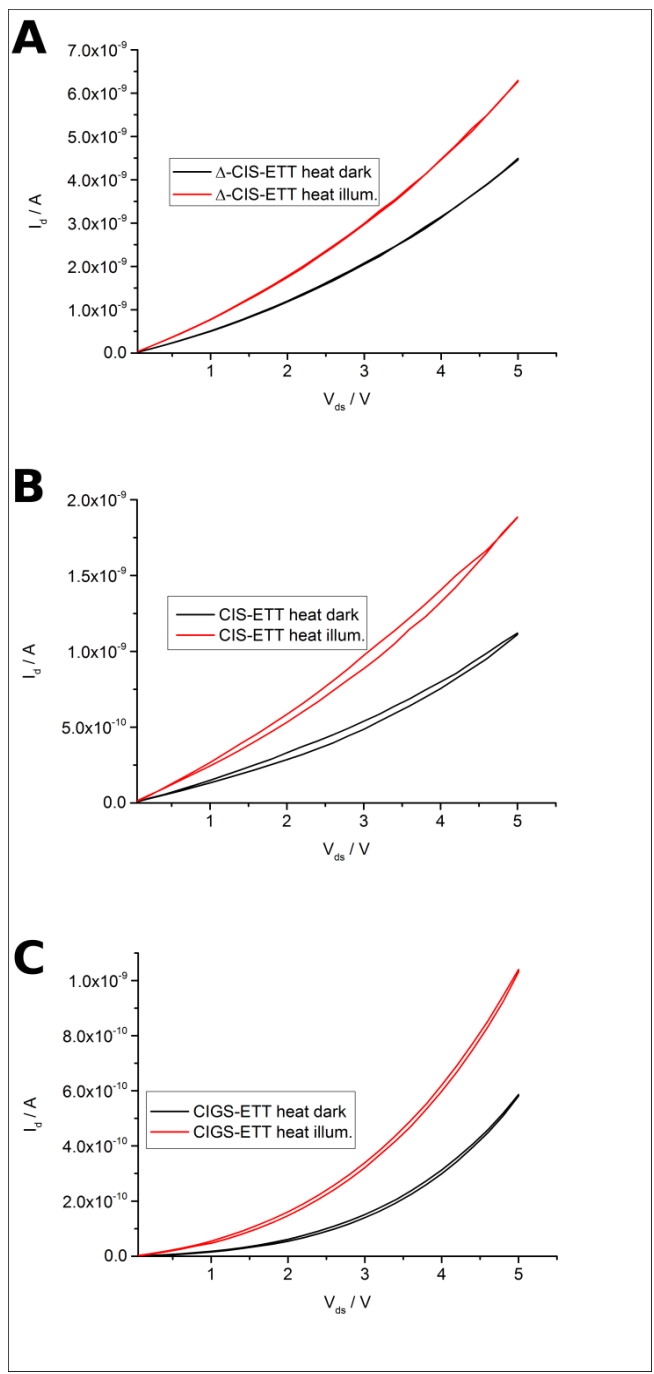

**Figure S16.** Photoresponse of the ligand-free chalcopyrite NC types (A) trigonal pyramidal CIS, (B) elongated CIS NCs, (C) CIGS NCs exhibiting 2 fold increase in photoresponse.